**The Quiescent Intracluster Medium in the Core of the Perseus Cluster**

The Hitomi collaboration

**Clusters of galaxies are the most massive gravitationally-bound objects in the Universe and are still forming. They are thus important probes[1] of cosmological parameters and a host of astrophysical processes. Knowledge of the dynamics of the pervasive hot gas, which dominates in mass over stars in a cluster, is a crucial missing ingredient. It can enable new insights into mechanical energy injection by the central supermassive black hole and the use of hydrostatic equilibrium for the determination of cluster masses. X-rays from the core of the Perseus cluster are emitted by the 50 million K diffuse hot plasma filling its gravitational potential well. The Active Galactic Nucleus of the central galaxy NGC1275 is pumping jetted energy into the surrounding intracluster medium, creating buoyant bubbles filled with relativistic plasma. These likely induce motions in the intracluster medium and heat the inner gas preventing runaway radiative cooling; a process known as Active Galactic Nucleus Feedback[2-6]. Here we report on Hitomi X-ray observations of the Perseus cluster core, which reveal a remarkably quiescent atmosphere where the gas has a line-of-sight velocity dispersion of 164±10 km/s in a region 30-60 kpc from the central nucleus. A gradient in the line-of-sight velocity of 150±70 km/s is found across the 60 kpc image of the cluster core. Turbulent pressure support in the gas is 4% or less of the thermodynamic pressure, with large scale shear at most doubling that estimate. We infer that total cluster masses determined from hydrostatic equilibrium in the central regions need little correction for turbulent pressure.**

The JAXA Hitomi X-ray Observatory[7] was launched on 2016 February 17 from Tanegashima, Japan. It carries the non-dispersive Soft X-ray Spectrometer SXS[8], which is a calorimeter cooled to 0.05K giving 4.9 eV FWHM (E/dE=1250 at 6 keV) Gaussian-shaped energy response over a 6 x 6 pixel array (total 3 x 3 arcmin). It operates over an energy range of 0.3-12 keV with X-rays focused by a mirror[9] with angular resolution of 1.2 arcmin (HPD). A gate valve was in place for early observations to minimize the risk of contamination from outgassing of the spacecraft. It includes a Be window that absorbs most X-rays below ~3 keV. The SXS can detect bulk and turbulent motions of the intracluster medium (ICM) by measuring Doppler shifts and broadening of the emission lines with unprecedented accuracy. It also allows the detection of weak emission lines or absorption features.

The SXS imaged a 60 x 60 kpc region in the Perseus cluster centered 1 arcmin to the NW of the nucleus for a total exposure time of 230 ks. The offset from the nucleus was due to the attitude control system not having then been calibrated. For this early observation, not all calibration procedures were available; in particular, we did not have contemporaneous calibration of the energy scale factors (gains) of the detector pixels. Gain variation over short time intervals was corrected using a separate calibration pixel illuminated by 5.9 keV Mn Kα photons from an $^{55}Fe$ X-ray source. Gain values were pinned to an absolute scale via extrapolation of a subsequent calibration of the whole array 10 days later using illumination by another $^{55}Fe$ source mounted on the filter wheel. (For more detail, see Methods.) We used a subset of the Perseus data closest to that calibration to derive the velocity map. For the line-width determination, we used the full dataset to minimize the statistical uncertainty, and applied a scale factor to force the Fe He-α complex from the cluster to have the same energy in all pixels. This minimizes the gain uncertainty in the determination of the velocity dispersion but also removes any true variations of the ICM line-of-sight velocity across the field.

A 5-8.5 keV spectrum of the full 3x3 arcmin field is shown in Fig 1. This spectrum shows a thermal continuum with line emission[10] from Cr, Mn, Fe and Ni. The strongest lines are from iron and consist of FeXXV He-α, β and γ complexes, together with FeXXVI Lyman α lines. The total number of counts in the He-α line is 21,726, of which about 16 counts are expected from residual instrumental background. The line complex is spread over about 75 eV and its major components include the resonance, intercombination and forbidden lines, all of which have been resolved.

We adopt a minimally model-dependent method for spectral fitting and represent the iron He-α, He-β and H-like Lyman α complexes in the spectrum with a set of Gaussians with free normalizations and energies fixed at redshifted laboratory energies in the case of He-like Fe and theory in the case of H-like Fe (Extended Data Table 1). Fig. 2 shows the profiles of these lines in a spectrum obtained from the outer region of the Perseus core which excludes the AGN and prominent inner bubbles (Fig. 3). In order to measure the line-of-sight velocity broadening (Gaussian $\sigma$), we fit the high signal-to-noise, Fe He-alpha line complex using 9 Gaussians associated with lines known from atomic physics and obtain 164 +/- 10 km/s (all uncertainties are quoted at the 90% confidence level). The widths of the 6.7008 keV resonance line and the 6.617 keV blend of faint satellite lines are allowed to be separate from the rest of the lines. The effect of the thermal broadening expected from the observed 4 keV plasma has been removed (alone it corresponds to 80 km/s). Conservative estimates of the uncertainty in energy resolution result in a systematic uncertainty range in the turbulent velocity of ±6 km/s. Uncertainties in plasma temperature add only a further ±2 km/s. The statistical scatter of our pixel self-calibration procedure results in an overestimate of the true broadening by not more than 3 km/s. The finite telescope angular resolution in the presence of a velocity gradient across the cluster results in a small artificial increase of the measured dispersion (see Methods) that is difficult to quantify at this stage.

The H-like Lyα complex alone (554 counts) yields a consistent velocity broadening of 160 ± 16 km/s. A search for spatial variations in velocity broadening using the He-α lines reveals that all 1 arcmin resolution bins give less than 200 km/s. With just a single pointing we cannot comment on how this picture translates to the wider cluster core.

The tightest previous constraint on the velocity dispersion of cluster gas was from the XMM-Newton Reflection Grating Spectrometer (RGS), giving[11,12] an upper limit of 235 km/s on the X-ray coolest gas (i.e. kT<3 keV) in the distant luminous cluster A1835. These measurements are available for only a few peaked clusters[13]; the angular size of Perseus and many other bright clusters is too large to derive meaningful velocity results from a slitless dispersive spectrometer like the RGS (the corresponding limit for Perseus[13] is 625 km/s). The Hitomi SXS achieves much higher accuracy on diffuse hot gas due to being non-dispersive.

We measure a slightly higher velocity broadening, 187±13 km/s, in the central region (Fig. 3a) that includes the bubbles and the nucleus. This region exhibits a strong power-law component from the Active Galactic Nucleus (AGN), several times brighter than the measurement[14] made in 2001 with XMM-Newton, consistent with the luminosity increase seen at other wavelengths. A fluorescent line from neutral Fe is present in the spectrum (Fig. 1), which can be emitted by the AGN or by the cold gas present in the cluster core[15]. The ICM has a slightly lower average temperature (3.8±0.1 keV) than the outer region (4.1±0.1 keV). Fitting lines with Gaussians, we measured the ratio of fluxes in He-$\alpha$ resonant to forbidden lines 2.48±0.16, which is lower than the expected value in optically thin plasma (for kT=3.8 keV, the current APEC[16] and SPEX[17] plasma models give ratios of 2.8 and 2.9-3.6) and suggests the presence of resonant scattering of photons[18]. Based on radiative transfer simulations[19] of resonant scattering in these lines, such resonance line suppression is in broad agreement with that expected for the measured low line widths and shear, providing independent indication of the low level of turbulence. Uncertainties in the current atomic data, as well as more complex structure along the line of sight and across the region complicate the interpretation of these results, which we defer to a future study that may indeed provide further information on ICM velocities.

A velocity map (Fig. 3b) has been produced from the absolute energies of the lines in the Fe He-α complex, using a subset of the data for which such a measurement was reliable, given the limited calibration (see Methods). We find a gradient in the line of sight velocities of about 150±70 km/s, from SE to NW of the SXS field of view. The velocity to the SE toward the nucleus is 48±17 (statistical) ± 50 km/s (systematic) km/s redshifted relative to NGC1275 (z=0.01756) and consistent with results from Suzaku CCD data[20]. Our statistical uncertainty on relative velocities is about 30 times better than that of Suzaku, although there is a systematic uncertainty on the absolute SXS velocities of about 50 km/s (see Methods).

NGC1275 hosts a giant (80 kpc wide) molecular nebula seen in CO and Hα of total cold gas mass several $10^{10}$ $M_{\odot}$, which dominate the total gas mass out to 15 kpc radius. The velocities of that gas[21,22] are consistent with the trend of the SXS bulk shear, suggesting that the molecular gas moves together with the hot plasma. (More details of the X-ray spectra and imaged region are given in Extended Data Figs 1-8.)

The large-scale bulk shear over the observed 60 kpc field is of comparable amplitude to the small-scale velocity dispersion that we derive for the outer region. The dispersion can be due to gas flows around the rising bubble at the centre of the field[23,24], a velocity gradient in the cold front[25] contained in this region, sound waves[26,27], turbulence[28] or galaxy motions[29]. The large-scale shear could be due to the buoyant AGN bubbles, or sloshing motions of gas in the cluster core that give rise to the cold front[25].

If the observed dispersion is interpreted as turbulence driven on scales comparable with the size of the largest bubbles in the field (~20-30 kpc), it is in agreement with the level inferred[28] from X-ray surface brightness fluctuations. In this case, our measured velocity dispersion suggests that turbulent dissipation of kinetic energy would be sufficient to offset radiative cooling. However, assuming isotropic turbulence, the ratio of turbulent pressure to thermal pressure in the ICM is low at 4%. Such low-velocity turbulence cannot spread far (<10 kpc) across the cooling core during the fraction (4%) of the cooling time in which it must be replenished, so the above mechanism requires that turbulence be generated in situ throughout the core. Another process is needed to transport the energy from the bubbling region. The observed level of turbulence is also sufficient to sustain the population of ultrarelativistic electrons giving rise to the radio synchrotron mini-halo observed in the Perseus core[30].

A low level of turbulent pressure and bulk shear, in a region continuously stirred by a central AGN and gas sloshing, is surprising and may imply that ICM turbulence is difficult to generate and/or easy to damp. If true throughout the cluster, this is encouraging for total mass measurements, which depend on knowledge of all forms of pressure support, and for cluster cosmology which depends on accurate masses.

The Hitomi spacecraft lost its ground contact on March 26, 2016, and later the recovery operation by JAXA was discontinued.

1. Allen, S.W., Evrard, A.E. & Mantz, A.B. Cosmological Parameters from Observations of Galaxy Clusters. *Ann.Rev.Astron.Astrophys.* **49,** 409-470 (2011)

2. Boehringer, H., Voges, W., Fabian, A. C., Edge, A. C. & Neumann, D. M. A ROSAT HRI study of the interaction of the X-ray-emitting gas and radio lobes of NGC 1275. *Mon. Not. R. Astron. Soc.* **264,** L25-L28 (1993)

3. Churazov, E., Forman, W., Jones, C. & Böhringer, H., Asymmetric, arc minute scale structures around NGC 1275*. Astron. Astrophys.* **356**, 788-794 (2000)

4. McNamara, B.R. et al. Chandra X-Ray Observations of the Hydra A Cluster: An Interaction between the Radio Source and the X-Ray-emitting Gas. *Astrophys.J.* **534**, L135-L138 (2000)

5. Fabian, A. C. et al. Chandra imaging of the complex X-ray core of the Perseus cluster. *Mon. Not. R. Astron. Soc*. **318,** L65-L68 (2000)

6. Fabian, A. C. *Ann. Rev. Astron. Astrophys*. **50,** 455-489 (2012)

7. Tadayuki, T. *et al*. The ASTRO-H X-ray astronomy satellite. *Proc. SPIE* **9144**, *Space Telescopes and Instrumentation 2014: Ultraviolet to Gamma Ray*, 914425 (2014)

8. Mitsuda, K. *et al*. Soft x-ray spectrometer (SXS): the high-resolution cryogenic spectrometer onboard ASTRO-H. *Proc. SPIE* **9144**, *Space Telescopes and Instrumentation 2014: Ultraviolet to Gamma Ray,* 91442A (2014)

9. Soong, Y. *et al.* ASTRO-H Soft X-ray Telescope (SXT). *Proc. SPIE* **9144**, *Space Telescopes and Instrumentation 2014: Ultraviolet to Gamma Ray*, 914428 (2014)

10. Tamura, T. *et al.* X-ray Spectroscopy of the Core of the Perseus Cluster with Suzaku: Elemental Abundances in the Intracluster Medium. *Astrophys. J.* **705,** L62-L66 (2009)

11. Sanders, J. S., Fabian, A. C., Smith, R. K. & Peterson, J. R. A direct limit on the turbulent velocity of the intracluster medium in the core of Abell 1835 from XMM-Newton. *Mon. Not. R. Astron. Soc*. **402** L11-L15 (2010)

12. Sanders, J. S., Fabian, A. C., & Smith, R. K. Constraints on turbulent velocity broadening for a sample of clusters, groups and elliptical galaxies using XMM-Newton. *Mon. Not. R. Astron. Soc*. **410** 1797-1812 (2011)

13. Pinto, C. et al. Chemical Enrichment RGS cluster Sample (CHEERS): Constraints on turbulence. *Astron.Astrophys.* **575**, 38 (2015)

14. Churazov, E., Forman, W., Jones, C. & Böhringer, H. XMM-Newton Observations of the Perseus Cluster. I. The Temperature and Surface Brightness Structure. Astrophys. J. **590**, 225-237 (2003)

15. Churazov, E., Sunyaev R., Gilfanov, M., Forman, W., Jones, C. The 6.4-keV fluorescent iron line from cluster cooling flows. *Mon. Not. R. Astron. Soc*. **297**,
1274-1278 (1998)

16. Foster, A. R., Li, J., Smith, R.K. & Brickhouse, N. S. Updated Atomic Data and Calculations for X-Ray Spectroscopy. *Astrophys.J.*, **756**, 128 (2012)

17. Kaastra, J. S., Mewe, R. & Nieuwenhuijzen, H., SPEX: a new code for spectral analysis of X & UV spectra, *11th Colloquium on UV and X-ray Spectroscopy of Astrophysical and Laboratory Plasmas*, p. 411 - 414, 1996

18. Gilfanov, M. R., Sunyaev, R.A. & Churazov, E. *Sov.Astron.Lett*. **13,** 3-7 (1987)

19. Zhuravleva, I. et al. Resonant scattering in the Perseus Cluster: spectral model for constraining gas motions with Astro-H. *Mon. Not. R. Astron. Soc*. **435,** 3111 (2013)

20. Tamura, T. et al. Gas Bulk Motion in the Perseus Cluster Measured with Suzaku Astrophys. J. **782** 38 (2014)

21. Salome, P. et al. A very extended molecular web around NGC 1275. *Astron.Astrophys*. **531**, 85 (2011)

22. Hatch, N.A., Crawford, C.S., Johnstone, R.M. & Fabian, A.C. On the origin and excitation of the extended nebula surrounding NGC1275. *Mon. Not. R. Astron. Soc.* **367**, 433-448 (2006)

23. Bruggen, M., Hoeft, M., Ruszkowski, M. X-Ray Line Tomography of AGN-induced Motion in Clusters of Galaxies. Astrophys.J, 628, 153-159 (2005)

24. Heinz, S., Bruggen, M. & Morsony, B. Prospects of High-Resolution X-ray Spectroscopy for Active Galactic Nucleus Feedback in Galaxy Clusters. *Astrophys.J.* **708**, 462-468 (2010)

25. Markevitch, M., Vikhlinin, A., Shocks and cold fronts in galaxy clusters. *Phys.Rep.* **443**, 1-53 (2007)

26. Fabian, A.C. et al. A deep Chandra observation of the Perseus cluster: shocks and ripples. *Mon. Not. R. Astron. Soc.* **344**, L43-L47 (2003)

27. Ruszkowski, M., Brüggen, M., Begelman, M. C. Cluster Heating by Viscous Dissipation of Sound Waves. *Astrophys.J*, **611**, 158-163 (2004)

28. Zhuravleva, I. et al. Turbulent heating in galaxy clusters brightest in X-rays. *Nature* **515,** 85-87 (2014)

29. Gu, L. et al. Probing of the Interactions between the Hot Plasmas and Galaxies in Clusters from z = 0.1 to 0.9. *Astrophys.J*. **767**, 157 (2013)

30. ZuHone, J.A., Markevitch, M. Brunetti, G. & Giacintucci, S. Turbulence and Radio Mini-halos in the Sloshing Cores of Galaxy Clusters. *Astrophys.J*. **762**, 78 (2013)

**Acknowledgements**:
We acknowledge all the JAXA members who have contributed to the ASTRO-H (Hitomi) project. All U.S. members gratefully acknowledge support through the NASA Science Mission Directorate. Stanford and SLAC members acknowledge support via DoE contract to SLAC National Accelerator Laboratory DE-AC3-76SF00515 and NASA grant NNX15AM19G. Part of this work was performed under the auspices of the U.S. DoE by LLNL under Contract DE-AC52-07NA27344 and also supported by NASA grants to LLNL. Support from the European Space Agency is gratefully acknowledged. French members acknowledge support from CNES, the Centre National d'Etudes Spatiales. SRON is supported by NWO, the Netherlands Organization for Scientific Research. Swiss team acknowledges support of the Swiss Secretariat for Education, Research and Innovation SERI and ESA's PRODEX programme. The Canadian Space Agency is acknowledged for the support of Canadian members. We acknowledge support from JSPS/MEXT KAKENHI grant numbers 15H02070, 15K05107, 23340071, 26109506, 24103002, 25400236, 25800119, 25400237, 25287042, 24540229, 25105516, 23540280, 25400235, 25247028, 26800095, 25400231, 25247028, 26220703, 24105007, 23340055, 15H00773, 23000004 15H02090, 15K17610, 15H05438, 15H00785, and 24540232. H. Akamatsu acknowledges support of NWO via Veni grant. M. Axelsson acknowledges JSPS International Research Fellowship. C. Done acknowledges STFC funding under grant ST/L00075X/1. P. Gandhi acknowledges JAXA International Top Young Fellowship and UK Science and Technology Funding Council (STFC) grant ST/J003697/2. H. Russell, A.C. Fabian and C.Pinto acknowledge support from ERC Advanced Grant Feedback 340442. We thank contributions by many companies, including in particular, NEC, Mitsubishi Heavy Industries, Sumitomo Heavy Industries, and Japan Aviation Electronics Industry.
Finally, we acknowledge strong support from the following engineers.

JAXA/ISAS: Chris Baluta, Nobutaka Bando, Atsushi Harayama, Kazuyuki Hirose, Kosei Ishimura, Naoko Iwata, Taro Kawano, Shigeo Kawasaki, Kenji Minesugi, Chikara Natsukari, Hiroyuki Ogawa, Mina Ogawa, Masayuki Ohta, Tsuyoshi Okazaki, Shin-ichiro Sakai, Yasuko Shibano, Maki Shida, Takanobu Shimada, Atsushi Wada, Takahiro Yamada; JAXA/TKSC: Atsushi Okamoto, Yoichi Sato, Keisuke Shinozaki, Hiroyuki Sugita; Chubu U: Yoshiharu Namba; Ehime U: Keiji Ogi; Kochi U of Technology: Tatsuro Kosaka; Miyazaki U: Yusuke Nishioka; Nagoya U: Housei Nagano; NASA/GSFC: Thomas Bialas, Kevin Boyce, Edgar Canavan, Michael DiPirro, Mark Kimball, Candace Masters, Daniel Mcguinness, Joseph Miko, Theodore Muench, James Pontius, Peter Shirron, Cynthia Simmons, Gary Sneiderman, Tomomi Watanabe; Noqsi Aerospace Ltd: John Doty; Stanford U/KIPAC: Makoto Asai, Kirk Gilmore; ESA (Netherlands): Chris Jewell; SRON: Daniel Haas, Martin Frericks, Philippe Laubert, Paul Lowes; U of Geneva: Philipp Azzarello; CSA: Alex Koujelev, Franco Moroso.

**Author contributions**

The science goals of Hitomi (known as ASTRO-H before launch) were discussed and developed over more than 10 years by the ASTRO-H Science Working Group (SWG), all members of which are authors of this manuscript. All the instruments were prepared by joint efforts of the team. Calibration of the Perseus dataset was carried out by members of the SXS team. Data analysis and manuscript preparation were carried out by a small subgroup of authors appointed by the SWG, based on the extensive discussion made in the white paper produced by all SWG members. The manuscript was subject to an internal collaboration-wide review process. All authors reviewed and approved the final version of the manuscript.

Correspondence to A.C.Fabian, Institute of Astronomy, Madingley Road, Cambridge CB3 0HA, UK (acf@ast.cam.ac.uk)

Author list:

**Hitomi Collaboration** Felix Aharonian[1,2] , Hiroki Akamatsu[3] , Fumie Akimoto[4], Steven W. Allen[5,6,7], Naohisa Anabuki[8], Lorella Angelini[9], Keith Arnaud[9,10], Marc Audard[11], Hisamitsu Awaki[12], Magnus Axelsson[13], Aya Bamba[14], Marshall Bautz[15], Roger Blandford[5,6,7], Laura Brenneman[16], Gregory V. Brown[17], Esra Bulbul[15], Edward Cackett[18], Maria Chernyakova[1], Meng Chiao[9], Paolo Coppi[19], Elisa Costantini[3], Jelle de Plaa[3], Jan-Willem den Herder[3], Chris Done[20], Tadayasu Dotani[21], Ken Ebisawa[21], Megan Eckart[9], Teruaki Enoto[22,23], Yuichiro Ezoe[13], Andrew Fabian[18], Carlo Ferrigno[11], Adam Foster[16], Ryuichi Fujimoto[24], Yasushi Fukazawa[25], Akihiro Furuzawa[4], Massimiliano Galeazzi[26], Luigi Gallo[27], Poshak Gandhi[28], Margherita Giustini[3], Andrea Goldwurm[29], Liyi Gu[3], Matteo Guainazzi[21,30], Yoshito Haba[31], Kouichi Hagino[21], Kenji Hamaguchi[9,32], Ilana Harrus[9,32], Isamu Hatsukade[33], Katsuhiro Hayashi[21], Takayuki Hayashi[4], Kiyoshi Hayashida[8], Junko Hiraga[34], Ann Hornschemeier[9], Akio Hoshino[35], John Hughes[36], Ryo Iizuka[21], Hajime Inoue[21], Yoshiyuki Inoue[21], Kazunori Ishibashi[4], Manabu Ishida[21], Kumi Ishikawa[37], Yoshitaka Ishisaki[13], Masayuki Itoh[38], Naoko Iyomoto[39], Jelle Kaastra[3], Timothy Kallman[9], Tuneyoshi Kamae[5], Erin Kara[10], Jun Kataoka[40], Satoru Katsuda[41], Junichiro Katsuta[25], Madoka Kawaharada[42], Nobuyuki Kawai[43], Richard Kelley[9], Dmitry Khangulyan[35], Caroline Kilbourne[9], Ashley King[5,6], Takao Kitaguchi[25], Shunji Kitamoto[35], Tetsu Kitayama[44], Takayoshi Kohmura[45], Motohide Kokubun[21], Shu Koyama[21], Katsuji Koyama[46], Peter Kretschmar[30], Hans Krimm[9,47], Aya Kubota[48], Hideyo Kunieda[4], Philippe Laurent[29], François Lebrun[29], Shiu-Hang Lee[21], Maurice Leutenegger[9], Olivier Limousin[29], Michael Loewenstein[9,10], Knox S. Long[49], David Lumb[50], Grzegorz Madejski[5,7], Yoshitomo Maeda[21], Daniel Maier[29], Kazuo Makishima[51], Maxim Markevitch[9], Hironori Matsumoto[52], Kyoko Matsushita[53], Dan McCammon[54], Brian McNamara[55], Missagh Mehdipour[3], Eric Miller[15], Jon Miller[56], Shin Mineshige[22], Kazuhisa Mitsuda[21], Ikuyuki Mitsuishi[4], Takuya Miyazawa[4], Tsunefumi Mizuno[25], Hideyuki Mori[9], Koji Mori[33], Harvey Moseley[9], Koji Mukai[9,32], Hiroshi Murakami[57], Toshio Murakami[24], Richard Mushotzky[10], Ryo Nagino[8], Takao Nakagawa[21], Hiroshi Nakajima[8], Takeshi Nakamori[58], Toshio Nakano[37], Shinya Nakashima[21], Kazuhiro Nakazawa[14], Masayoshi Nobukawa[59], Hirofumi Noda[37], Masaharu Nomachi[60], Steve O' Dell[61], Hirokazu Odaka[21], Takaya Ohashi[13], Masanori Ohno[25], Takashi Okajima[9], Naomi Ota[62], Masanobu Ozaki[21], Frits Paerels[63], Stephane Paltani[11], Arvind Parmar[50], Robert Petre[9], Ciro Pinto[18], Martin Pohl[11], F. Scott Porter[9], Katja Pottschmidt[9,32], Brian Ramsey[61], Christopher Reynolds[10], Helen Russell[18], Samar Safi-Harb[64], Shinya Saito[35], Kazuhiro Sakai[9], Hiroaki Sameshima[21], Goro Sato[21], Kosuke Sato[53], Rie Sato[21], Makoto Sawada[65], Norbert Schartel[30], Peter Serlemitsos[9], Hiromi Seta[13], Megumi Shidatsu[51], Aurora Simionescu[21], Randall Smith[16], Yang Soong[9], Lukasz Stawarz[66], Yasuharu Sugawara[41], Satoshi Sugita[43], Andrew Szymkowiak[19], Hiroyasu Tajima[67], Hiromitsu Takahashi[25], Tadayuki Takahashi[21], Shin'ichiro Takeda[68], Yoh Takei[21], Toru Tamagawa[37], Keisuke Tamura[4], Takayuki Tamura[21], Takaaki Tanaka[46], Yasuo Tanaka[21], Yasuyuki Tanaka[25], Makoto Tashiro[69], Yuzuru Tawara[4],

Yukikatsu Terada[69], Yuichi Terashima[12], Francesco Tombesi[9], Hiroshi Tomida[21], Yohko Tsuboi[41], Masahiro Tsujimoto[21], Hiroshi Tsunemi[8], Takeshi Tsuru[46], Hiroyuki Uchida[46], Hideki Uchiyama[70], Yasunobu Uchiyama[35], Shutaro Ueda[21], Yoshihiro Ueda[22], Shiro Ueno[21], Shin'ichiro Uno[71], Meg Urry[19], Eugenio Ursino[26], Cor de Vries[3], Shin Watanabe[21], Norbert Werner[5,6], Daniel Wik[9,72], Dan Wilkins[27], Brian Williams[9], Shinya Yamada[13], Hiroya Yamaguchi[9], Kazutaka Yamaoka[4], Noriko Y. Yamasaki[21], Makoto Yamauchi[33], Shigeo Yamauchi[62], Tahir Yaqoob[32,9], Yoichi Yatsu[43], Daisuke Yonetoku[24], Atsumasa Yoshida[65], Takayuki Yuasa[37], Irina Zhuravleva[5,6], and Abderahmen Zoghbi[56]

[1]Astronomy and Astrophysics Section, Dublin Institute for Advanced Studies, Dublin 2, Ireland [2]National Research Nuclear University (MEPHI), 115409, Moscow, Russia [3]SRON Netherlands Institute for Space Research, Utrecht, The Netherlands [4]Department of Physics, Nagoya University, Aichi 464-8602, Japan [5]Kavli Institute for Particle Astrophysics and Cosmology, Stanford University, CA 94305, USA [6]Department of Physics, Stanford University, 382 Via Pueblo Mall, Stanford, CA 94305, USA [7]SLAC National Accelerator Laboratory, 2575 Sand Hill Road, Menlo Park, CA 94025, USA [8]Department of Earth and Space Science, Osaka University, Osaka 560-0043, Japan [9]NASA/Goddard Space Flight Center, MD 20771, USA [10]Department of Astronomy, University of Maryland, MD 20742, USA [11]Université de Genève, 1211 Genève 4, Switzerland [12]Department of Physics, Ehime University, Ehime 790-8577, Japan [13]Department of Physics, Tokyo Metropolitan University, Tokyo 192-0397, Japan [14]Department of Physics, University of Tokyo, Tokyo 113-0033, Japan [15]Kavli Institute for Astrophysics and Space Research, Massachusetts Institute of Technology, MA 02139, USA [16]Smithsonian Astrophysical Observatory, 60 Garden St., MS-4, Cambridge, MA 02138, USA [17]Lawrence Livermore National Laboratory, CA 94550, USA [18]Institute of Astronomy, Cambridge University, CB3 0HA, UK [19]Yale Center for Astronomy and Astrophysics, Yale University, CT 06520-8121, USA [20]Department of Physics, University of Durham, DH1 3LE, UK [21]Institute of Space and Astronautical Science (ISAS), Japan Aerospace Exploration Agency (JAXA), Kanagawa 252-5210, Japan [22]Department of Astronomy, Kyoto University, Kyoto 606-8502, Japan [23]The Hakubi Center for Advanced Research, Kyoto University, Kyoto 606-8302, Japan [24]Faculty of Mathematics and Physics, Kanazawa University, Ishikawa 920-1192, Japan [25]Department of Physical Science, Hiroshima University, Hiroshima 739-8526, Japan [26]Physics Department, University of Miami, FL 33124, USA [27]Department of Astronomy and Physics, Saint Mary's University, Halifax, NS B3H 3C3, Canada [28]Department of Physics and Astronomy, University of Southampton, Highfield, Southampton, SO17 1BJ, UK [29]IRFU/Service d'Astrophysique, CEA Saclay, 91191 Gif-sur-Yvette Cedex, France [30]European Space Agency (ESA), European Space Astronomy Centre (ESAC), Madrid, Spain [31]Department of Physics and Astronomy, Aichi University of Education, Aichi 448-8543, Japan [32]Department of Physics, University of Maryland, Baltimore County, 1000 Hilltop Circle, Baltimore, MD

21250, USA [33]Department of Applied Physics and Electronic Engineering, University of Miyazaki, Miyazaki 889-2192, Japan [34]Department of Physics, School of Science and Technology, Kwansei Gakuin University, 669-1337, Japan [35]Department of Physics, Rikkyo University, Tokyo 171-8501, Japan [36]Department of Physics and Astronomy, Rutgers University, NJ 08854-8019, USA [37]RIKEN Nishina Center, Saitama 351-0198, Japan [38]Faculty of Human Development, Kobe University, Hyogo 657-8501, Japan [39]Kyushu University, Fukuoka 819-0395, Japan [40]Research Institute for Science and Engineering, Waseda University, Tokyo 169-8555, Japan [41]Department of Physics, Chuo University, Tokyo 112-8551, Japan [42]Tsukuba Space Center (TKSC), Japan Aerospace Exploration Agency (JAXA), Ibaraki, 305-8505, Japan [43]Department of Physics, Tokyo Institute of Technology, Tokyo 152-8551, Japan [44]Department of Physics, Toho University, Chiba 274-8510, Japan [45]Department of Physics, Tokyo University of Science, Chiba 278-8510, Japan [46]Department of Physics, Kyoto University, Kyoto 606-8502, Japan [47]Universities Space Research Association, 7178 Columbia Gateway Dr., Columbia, MD 21046, USA [48]Department of Electronic Information Systems, Shibaura Institute of Technology, Saitama 337-8570, Japan [49]Space Telescope Science Institute, MD 21218, USA [50]European Space Agency (ESA), European Space Research and Technology Centre (ESTEC), 2200 AG Noordwijk, The Netherlands [51]RIKEN, Saitama 351-0198, Japan [52]Kobayashi-Maskawa Institute, Nagoya University, Aichi 464-8602, Japan [53]Department of Physics, Tokyo University of Science, Tokyo 162-8601, Japan [54]Department of Physics, University of Wisconsin, WI 53706, USA [55]University of Waterloo, Ontario N2L 3G1, Canada [56]Department of Astronomy, University of Michigan, MI 48109, USA [57]Department of Information Science, Faculty of Liberal Arts, Tohoku Gakuin University, Miyagi 981-3193, Japan [58]Department of Physics, Faculty of Science, Yamagata University, Yamagata 990-8560, Japan [59]Department of Teacher Training and School Education, Nara University of Education, Takabatake-cho, Nara 630-8528, Japan [60]Research Center for Nuclear Physics (Toyonaka), Osaka University, 1-1 Machikaneyama-machi, Toyonaka, Osaka 560-0043, Japan [61]NASA/Marshall Space Flight Center, AL 35812, USA [62]Department of Physics, Faculty of Science, Nara Women's University, Nara 630-8506, Japan [63]Department of Astronomy, Columbia University, NY 10027, USA [64]Department of Physics and Astronomy, University of Manitoba, MB R3T 2N2, Canada [65]Department of Physics and Mathematics, Aoyama Gakuin University, Kanagawa 252-5258, Japan [66]Astronomical Observatory, Jagiellonian University, 30-244, Poland [67] Institute of Space-Earth Environmental Research, Nagoya University, Aichi 464-8601, Japan [68]Advanced Medical Instrumentation Unit, Okinawa Institute of Science and Technology Graduate University (OIST), Okinawa 904-0495, Japan [69]Department of Physics, Saitama University, Saitama 338-8570, Japan [70]Science Education, Faculty of Education, Shizuoka University, Shizuoka 422-8529, Japan [71]Faculty of Health Sciences, Nihon Fukushi University, Aichi 475-0012, Japan [72]Department of Physics and Astronomy, Johns Hopkins University, MD 21218, USA

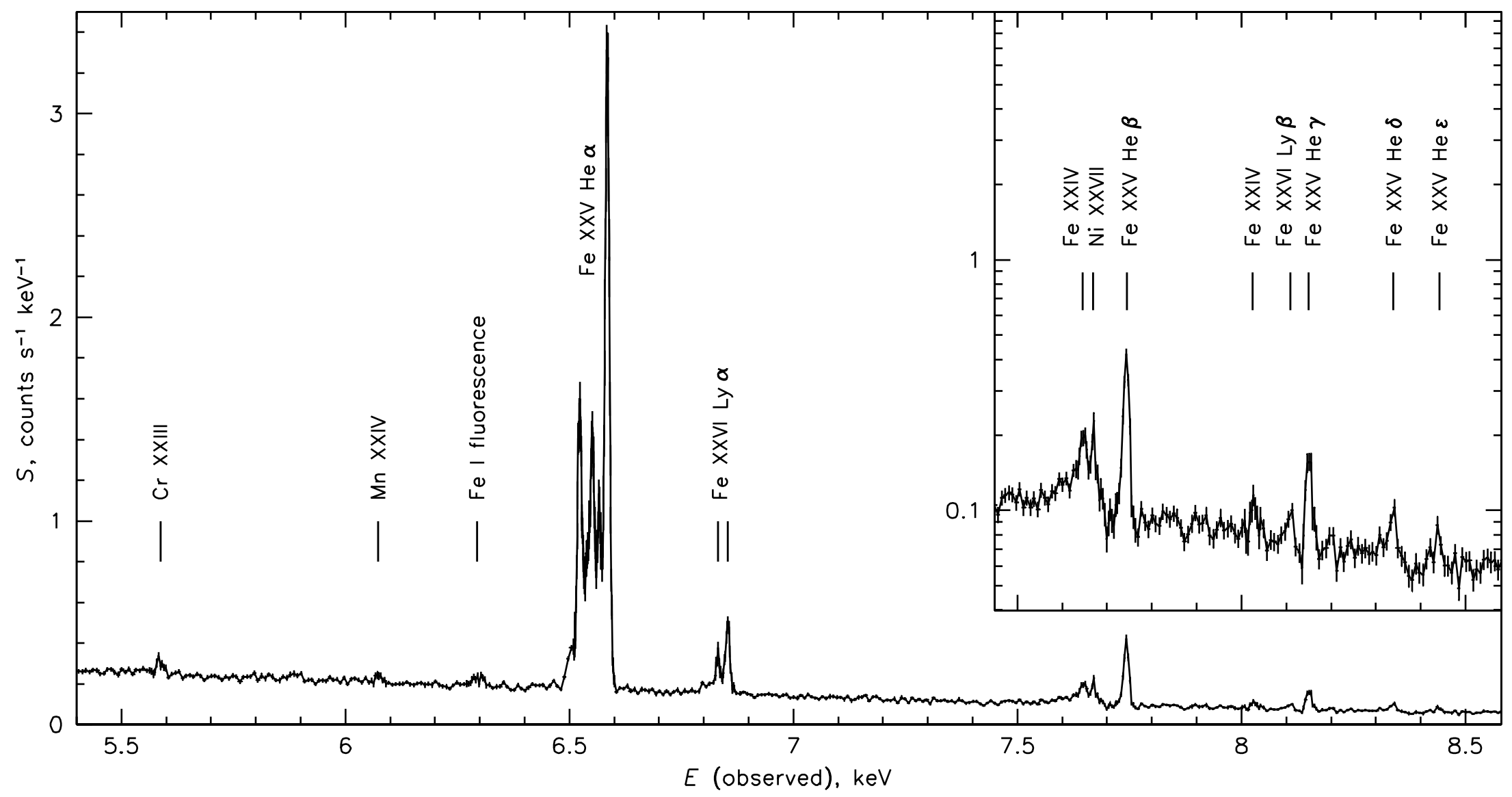

Fig 1 Full array spectrum of the Perseus cluster core obtained by the Hitomi observatory. The redshift of the Perseus cluster is 0.01756. The inset above 7.5 keV has a log scale which allows the weaker lines to be better seen.

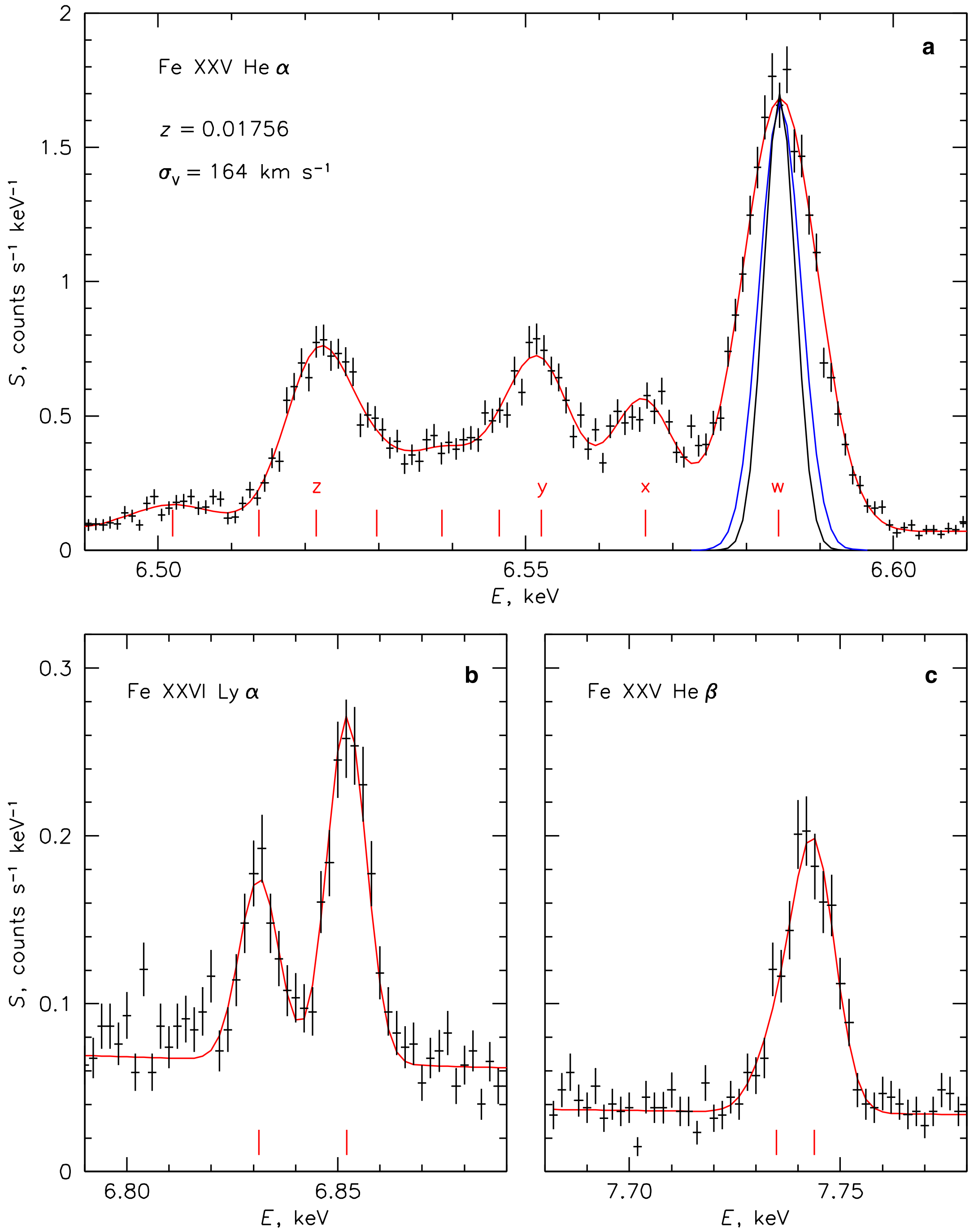

Fig. 2. Spectra of FeXXV He-α, XXVI Lyman α and XXV He-β from the outer region. Gaussian fits have been made to lines with energies (marked in red) from laboratory measurements in the case of He-like Fe XXV, and theory in the case of Fe XXVI (see Extended Data Table 1 for details) with the same velocity dispersion, except for the He-α resonant line which was allowed to have its own width. Instrumental broadening with (blue line) and without (black line) thermal broadening are indicated. The redshift is the cluster value to which the data were self-

calibrated using the He-α lines. The strongest resonance (w), intercombination (x,y) and forbidden (z) lines are indicated.

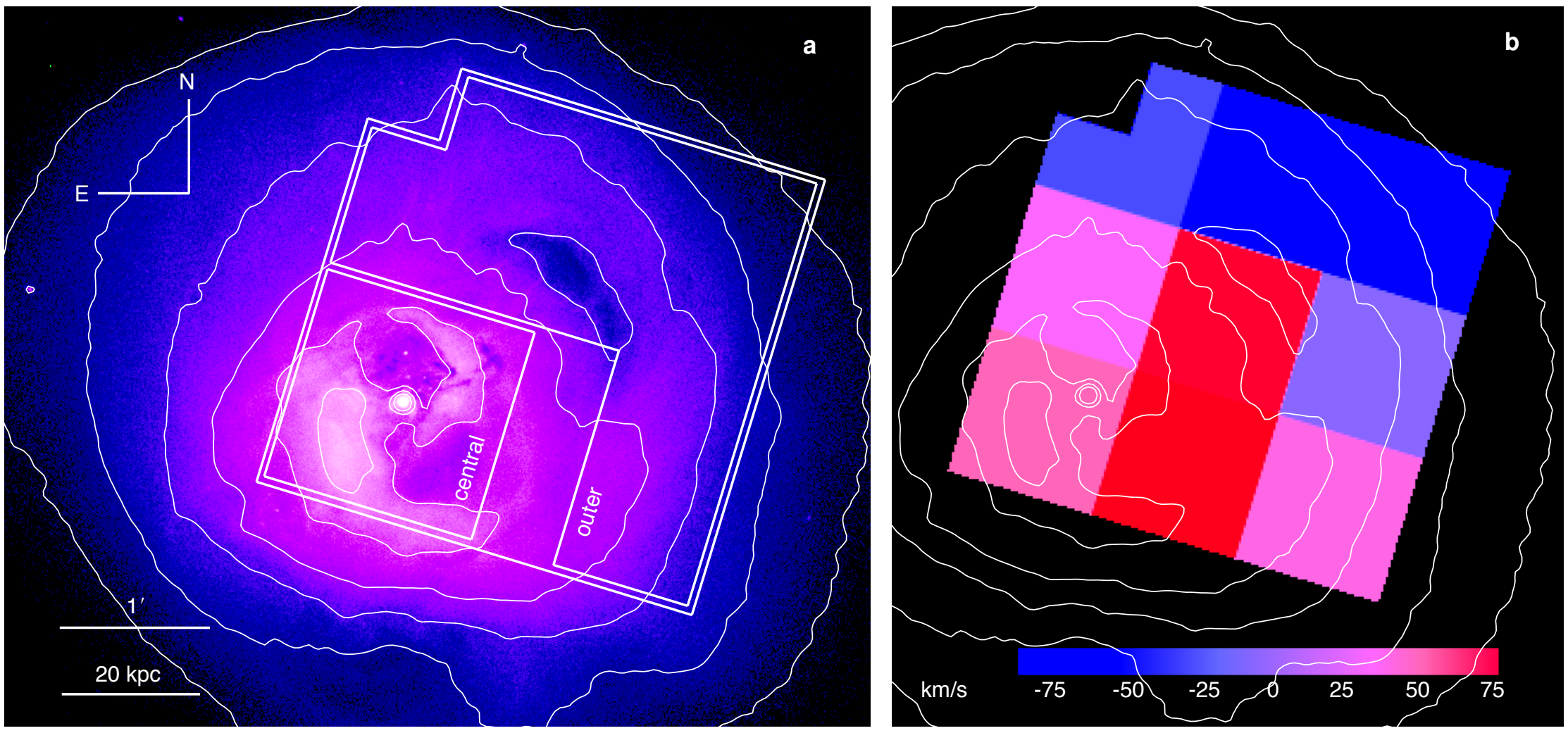

Fig. 3. The region of the Perseus cluster and velocity field viewed by the SXS.
a) The field of view of the SXS overlaid on a Chandra image. The nucleus of NGC1275 is seen as the white dot with inner bubbles N and S. A buoyant outer bubble lies NW of the centre of the field. A swirling cold front coincides with the second contour in from the outside. The central and outer regions are marked.

b) The bulk velocity field across the imaged region. Colors show the difference from the velocity of the central galaxy NGC 1275 (whose redshift is z=0.01756); positive difference means gas receding faster than the galaxy. The one arcmin pixels of the map correspond approximately to the angular resolution, but are not entirely independent (see Extended Data). The calibration uncertainty on velocities in individual pixels and in the overall baseline is 50 km/s (Δz=0.00017).

## METHODS

### Gain corrections and calibration

Gain scales for each pixel were measured in ground calibration using a series of fiducial x-ray lines at several detector heat sink temperatures (a single spectral energy reference is sufficient to determine the effective detector temperature and thus the appropriate gain curve to use). As the heat-sink temperature varies, the gain of each pixel tracks the gain change in the separate calibration pixel that is continuously illuminated by a dedicated $^{55}Fe$ source. However, time-varying differential thermal loading of the pixels changes their gains by different factors. Thus, use of the gain history of the calibration pixel alone can be insufficient to correct the gain scale of the main array.

The Perseus observation used for this work was performed in two parts, 7 days apart, during which the gain of the calibration pixel changed by 0.6%. Ten days after the last observation, a fiducial measurement for the full array was obtained with an on-board $^{55}Fe$ source mounted on a filter wheel. In order to relate this calibration to the two Perseus observations, a two-stage approach was used. A first correction factor was applied to all pixels using the gain history of the calibration pixel. Second, the differential pixel-pixel gain error was removed using the science observation itself. To do this, the two Perseus observations were subdivided, and the He-like Fe complex was fit for each pixel in each subset. The time dependent relative gain of each pixel (compared to the gain correction of the calibration pixel) was then linearly fit and extrapolated to the later full-array calibration. The full dataset was then corrected using this time-dependent gain function, and the fit errors were incorporated into the error analysis. To validate this approach, we compared the first observation, which required a substantial gain correction, to the second, for which the instrument was much closer to thermal equilibrium and thus required much less correction. In the first case, the bulk velocity uncertainties are dominated by the uncertainties in the gain correction, where in the second the uncertainties are dominated by the fit to the He-like Fe complex. The results for the two data sets agree for both bulk velocity and velocity dispersion, indicating that this is a robust approach. For the absolute velocity maps, we are presenting only the result from the second observation of the two used in this work, which requires the least correction and thus has the smallest uncertainty.

To derive the cluster absolute velocities, we applied the heliocentric correction, which was -26.4 km/s for the observation used for velocity mapping. The orbital motion of the satellite around Earth averages out. Our velocities are compared to the heliocentric velocity of NGC 1275 in Figs. 3 and E6.

An additional validation of our calibration comes from a weak background line in the whole-array spectrum from stray $^{55}$Fe X-rays, which, after the above procedure, is observed at the correct energy to ± 1.8 eV (equivalent to ±90 km/s). Though the line is not strong enough to verify the calibration of individual pixels (as there should be about 68 counts in this line, non-uniformly distributed across the array), it is a convincing check of the approach.

To determine velocity dispersion, we applied additional scale factors for each SXS pixel to match the apparent energies of the cluster Fe He-α complex in order to remove any residual gain errors at the relevant energy. This also removes the effect of true bulk shear. Pixels were then combined in physically relevant regions to minimize statistical uncertainties.

We have presumed a fixed energy resolution of 5.0 eV FWHM in all the analysis. Comparing the line widths in the first and second parts of the observation in order to estimate the broadening from residual gain drift, and accounting for the variation in resolution of the calibration pixel in time over the observation and during the later calibration of the array, we estimate that the composite resolution of the array and of the separately analyzed central and outer regions is bounded with high confidence between 4.5 eV and 5.5 eV FWHM. This 10% uncertainty in instrumental broadening produces a much smaller fractional uncertainty in velocity broadening because the instrumental broadening is roughly half as large as the astronomical broadening, and adds in quadrature with it.

The error from energy-matching the different pixels in a region is smaller than this because of the small statistical errors in the determination of the scale factor at the Fe He-alpha complex (in an outer pixel, equivalent to 30 km/s at 90% confidence). Adding the spectra of multiple pixels with the same velocity uncertainty will add 30 km/s of noise in quadrature with the measured broadening, producing an overestimate by no more than 3 km/s.

Our velocity dispersion measurements exclude velocity variations across the field on scales 20 kpc and above because of the above self-calibration procedure, but integrate over all scales along the line of sight (weighted by X-ray emissivity, which essentially limits integration to the cluster core). Any comparison with simulations will have to take these into account.

**Effects of angular resolution**

The telescope point spread function (PSF) has a 1.2' half-power diameter (HPD) as measured during ground calibration. This means that regions used for spectral extraction get photons not only from the corresponding cluster regions in the sky, but also from the surrounding regions. The PSF image is shown in right panel of Fig. E5, centered on the SXS pixel that contains the cluster peak. Comparing the PSF with the middle panel showing the image in Fe He-a line (which comes mostly from the gas, as opposed to the central AGN), we see that the cluster diffuse emission is resolved. However, small regions in the detector, such as the 1'x1' regions of the velocity map shown in Figs. 3b and E6, are significantly correlated. The fraction of the emission that originates in a given 1' cluster region and ends up in the corresponding 1' detector region is 36-37%, with the rest spreading over the surrounding regions. For example, for the region marked -60 in Fig. E6, the scattered contribution from the neighboring region marked 78 is 23% of the flux that originates in region -60 itself; the contribution from -60 into 78 is a similar 22% of the flux that originates and stays in 78. Regions adjacent to the brightness peak (in region 48) are most affected - region 94 has a ratio of photons scattered in from 48 to its own photons of 27%. This means that the true l.o.s. velocity gradients on a 1' scale have to be steeper than what we measure, but not by much. Scattered flux from an adjacent region with a large velocity difference (e.g., from region 78 to region -60) should contribute lines at a different velocity in the spectrum, but such contributions would drown in the observed l.o.s. velocity dispersion of >160 km/s. Correction of the PSF effects is left for future work. Note that the limited gain calibration results in pixel-to-pixel uncertainty of 50 km/s on the absolute velocities.

The PSF scattering also has a subtle effect of inflating our measured value of velocity dispersion. While the self-calibration procedure that aligns the Fe He-α lines in each pixel (as described above) removes most of the velocity gradient contribution from the measured velocity dispersion, it does so after the PSF scattering has occurred and mixed the photons from regions with different l.o.s velocities, so that contribution remains.

**Pointing**

For this early observation, accurate pointing direction of the spacecraft was not available. We therefore assumed that the observed brightness peak in the SXS image is the AGN in NGC1275. The resulting uncertainty of the sky coordinates should be less than 15". The peak of the source determined in short time intervals revealed a small drift of the source in the detector image, within the above coordinate uncertainty. It causes image smearing that is insignificant compared to the PSF scattering effect.

**Additional References**

31. Grevesse, N. & Sauval, A.J. Standard Solar Composition. *SpaceSci.Rev*. **81**, 161-174 (1998)

32. Conselice, C., Gallagher, J.G & Wyse, R.G., On the Nature of the NGC 1275 System. *Astron.J.*, **122**, 2281-2300 (2001)

33. Fabian, A.C. *et al*. A wide Chandra view of the core of the Perseus cluster. *Mon. Not. R. Astron. Soc*. **418,** 2154-2164 (2011)

34. Beiersdorfer, P. et al. High-resolution measurements, line identification, and spectral modeling of K-alpha transitions in Fe XVIII-Fe XXV. *Astrophys.J*. **409**, 846-859 (1993)

35. Smith, A.J. et al. Kβ spectra of heliumlike iron from tokomak-fusion-test-reactor plasmas. *Phys.Rev.A* **47***,* 3073-3079 (1993)

36. Johnson, W.R. & Soff, G. The Lamb Shift in Hydrogen-like atoms, $1 \leq Z \leq 110$. Atomic Data and Nuclear Data Tables **33,** 405-466 (1985)

37. Ferruit, P., Binette, L., Sutherland, R.S., Pencontal, E., Tiger observations of the low and high velocity components of NGC 1275, *New Astron.* **2**, 345-363 (1997)

## EXTENDED DATA

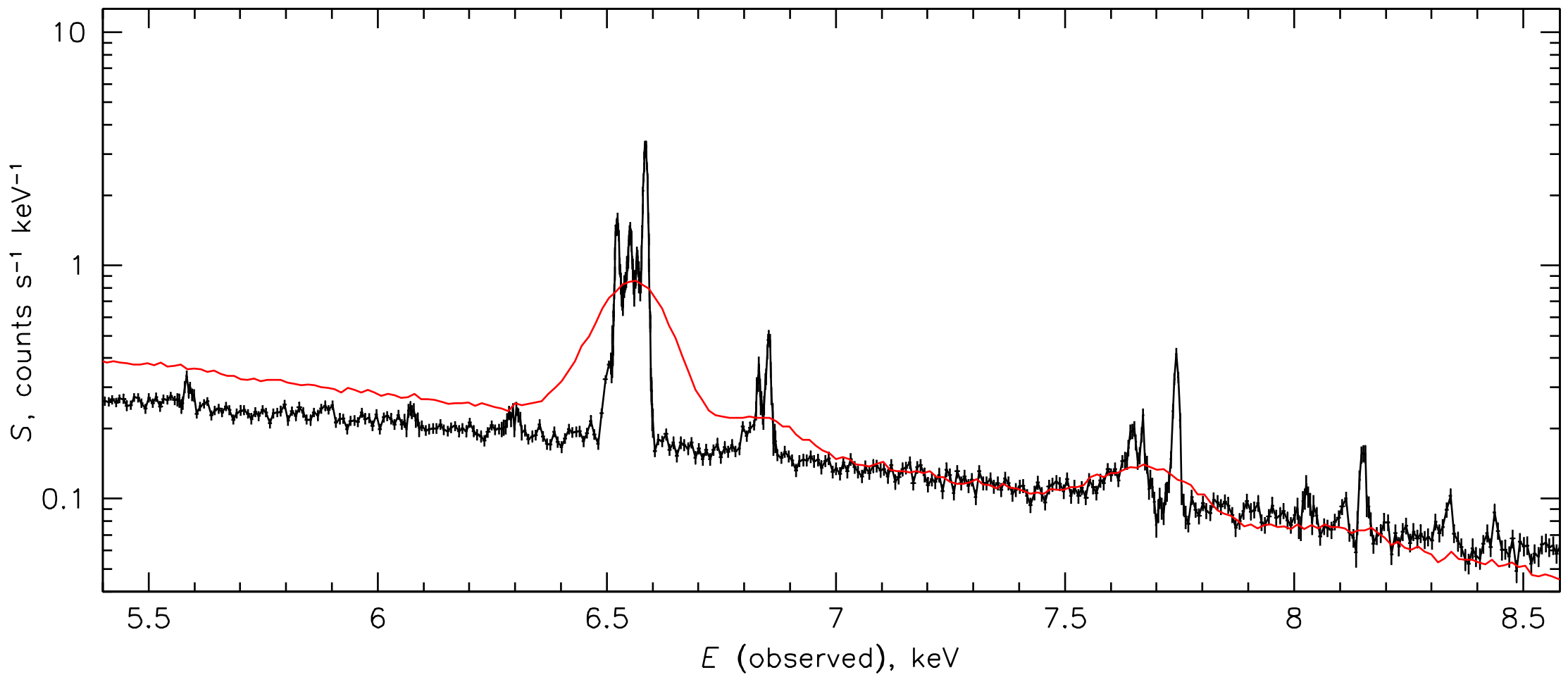

Fig E1. SXS spectrum of the full field overlaid with a CCD spectrum of the same region. The CCD is the Suzaku XIS (red line); the difference in the continuum slope is due to differences in the effective areas of the instruments.

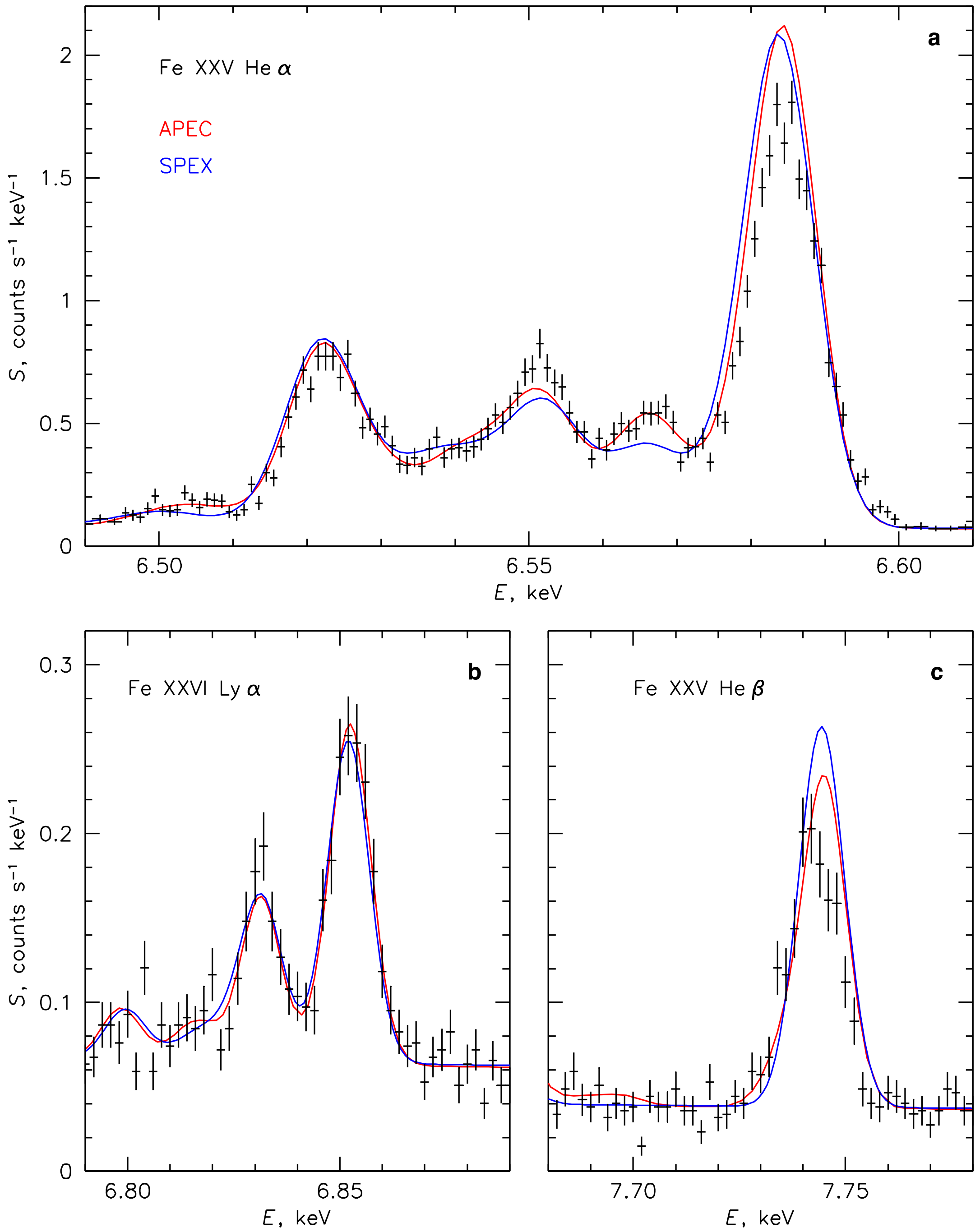

Fig E2. The iron line complexes from the outer region compared with best-fit models. These have been obtained from various emission line databases typically used in the literature. The spectra were modelled as a single temperature, optically thin plasma in collisional ionisation equilibrium using either APEC/ATOMDB 3.0.3 (ref 16; red) or SPEX 3.0 (ref. 17; blue). We determined the best-fit model by fitting the Hitomi spectrum from the outer 23 pixels in the energy range 6.4-8 keV, excluding the Fe He-α resonance line and Ni He-α line complex. We obtain consistent best-fit parameters, with both APEC and SPEX

predicting a temperature of 4.1±0.1 keV. The iron to hydrogen abundances are 0.62±0.02 from APEC and 0.74±0.02 from SPEX, relative to Solar values[31]. The line broadening obtained from APEC, 146±7 km/s, is smaller than the best-fit SPEX value of 171±7, although both values are consistent with the line broadening obtained by fitting a set of Gaussians (the result presented in the main body of the paper). Apart from the Fe He-α w line affected by resonance scattering, both emission line models presented here currently have difficulty reproducing the measured Fe He-α intercombination lines as well as the exact position of the Fe He-β line. This motivates the model-independent approach to determining the line widths adopted in the manuscript.

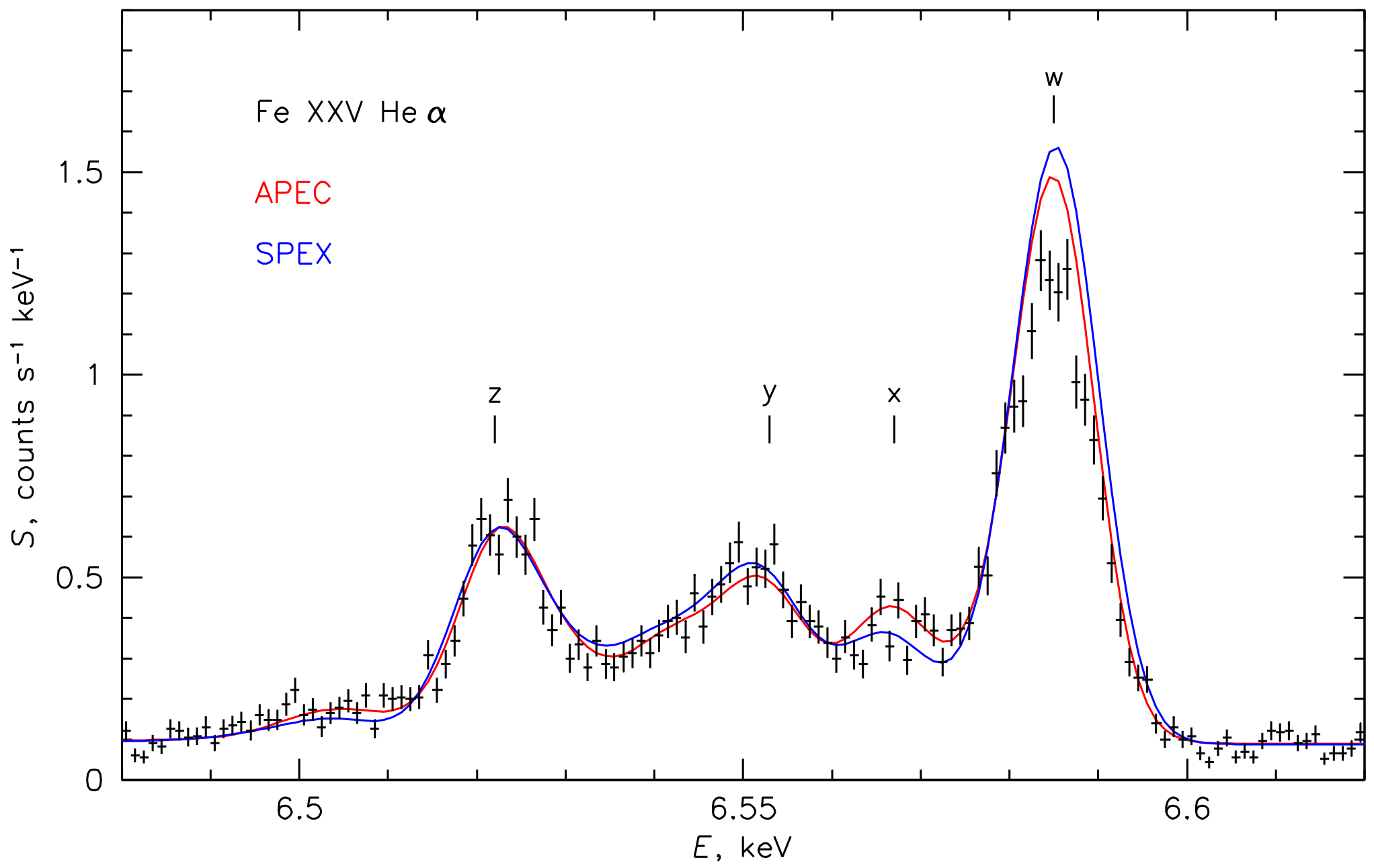

Fig E3. The Fe He-α line complex from the central region around the AGN is displayed. The 5.0-8.5 keV spectrum was modelled with an isothermal optically thin plasma in collisional ionisation equilibrium using either APEC/ATOMDB 3.0.3 (red) or SPEX 3.0 (blue), with an additional power-law component accounting for emission from the central AGN. During the fit we have excluded the Fe He-α resonance line because this can be affected by resonant scattering of photons by the intracluster gas in the line of sight. The two spectral codes provide similar results with an average temperature of 3.8 ± 0.1 keV and metallicity consistent with the solar value. We obtain a velocity broadening of 156±12 km/s from APEC and 178±9 km/s from SPEX.

Both models suggest that the resonant line has been suppressed in the central region.

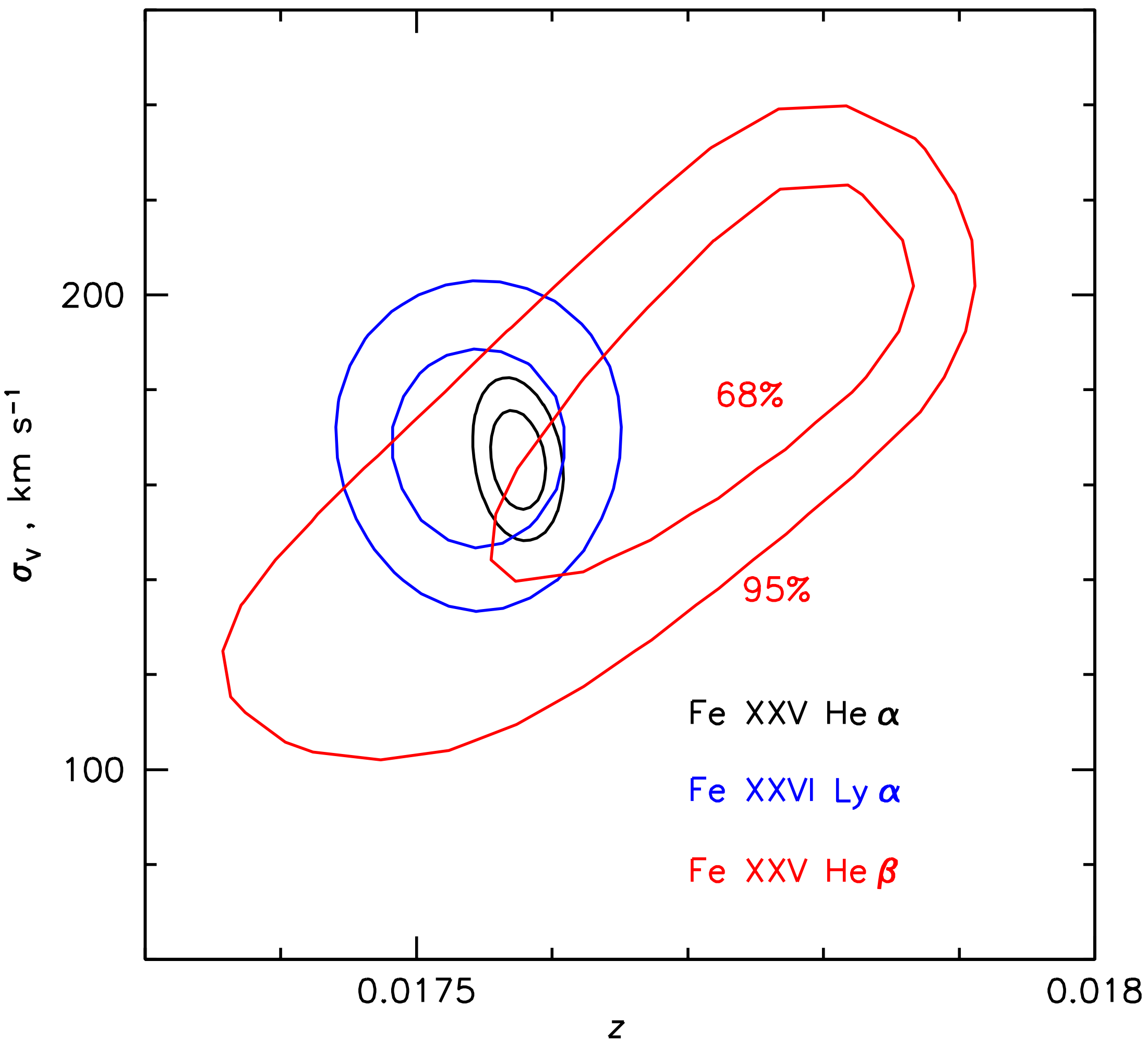

Fig E4. Confidence contours for joint fits of redshift and velocity broadening are compared. The three line complexes have been fitted independently. The contours are plotted at $\chi^2_{min}$ + 2.3 (68%, two parameters) and +6.17 (95%). The three fits give consistent redshifts (with the one to which the data were self-calibrated) and broadening.

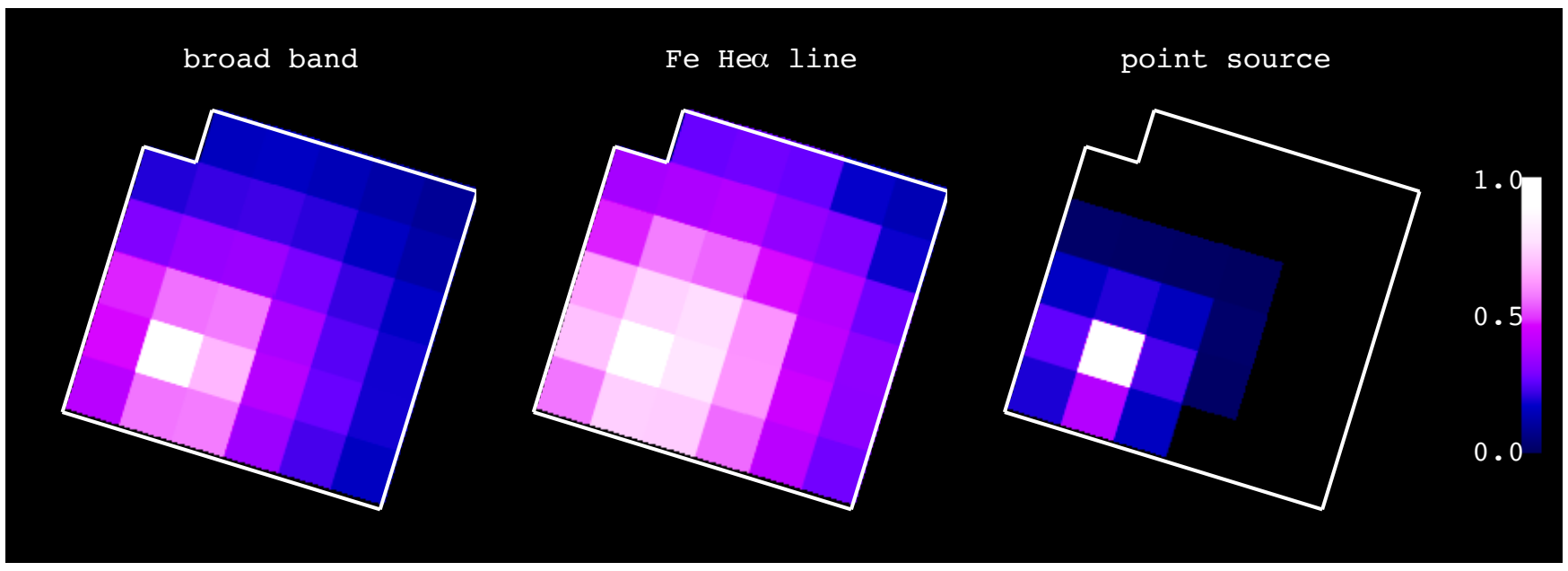

Fig E5. The spatial response of the SXS array is illustrated. The total broadband counts seen across the detector array (left), FeXXV He-α line counts (centre) that come mostly from the diffuse cluster plasma, and a model response of a point source centred in the pixel coincident with the nucleus of NGC1275 (right) are compared. Brightness is normalized to the same peak value.

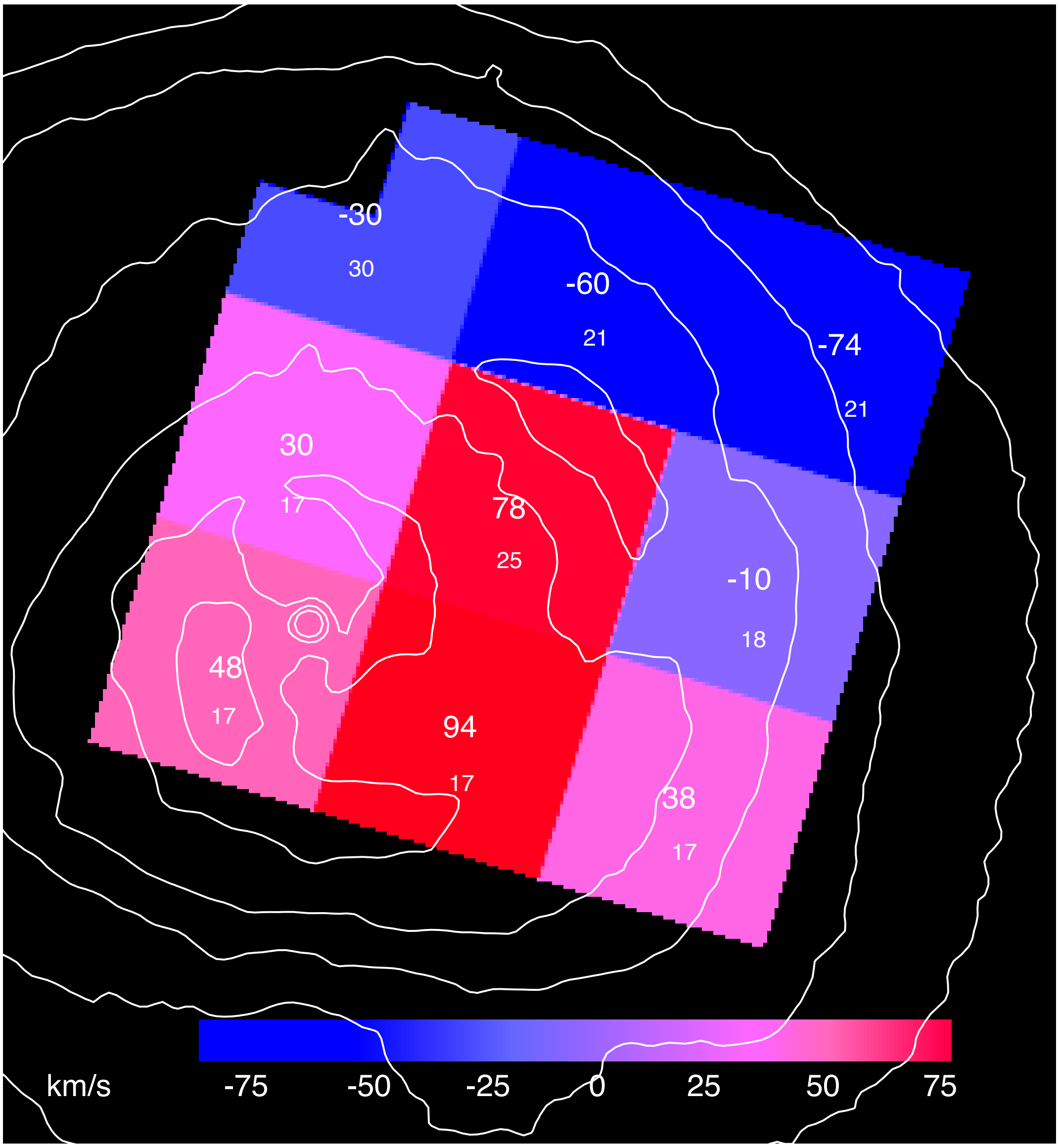

Fig E6. The line-of-sight gas velocities are overlaid on a deep Chandra image^33. The contours increase by a factor of 1.5. The 90% errors in the figure are statistical only; our estimate of the calibration uncertainty in individual pixels is 50 km/s. Heliocentric correction has been applied. Velocities are shown relative to that of NGC1275, whose redshift is z=0.01756^37

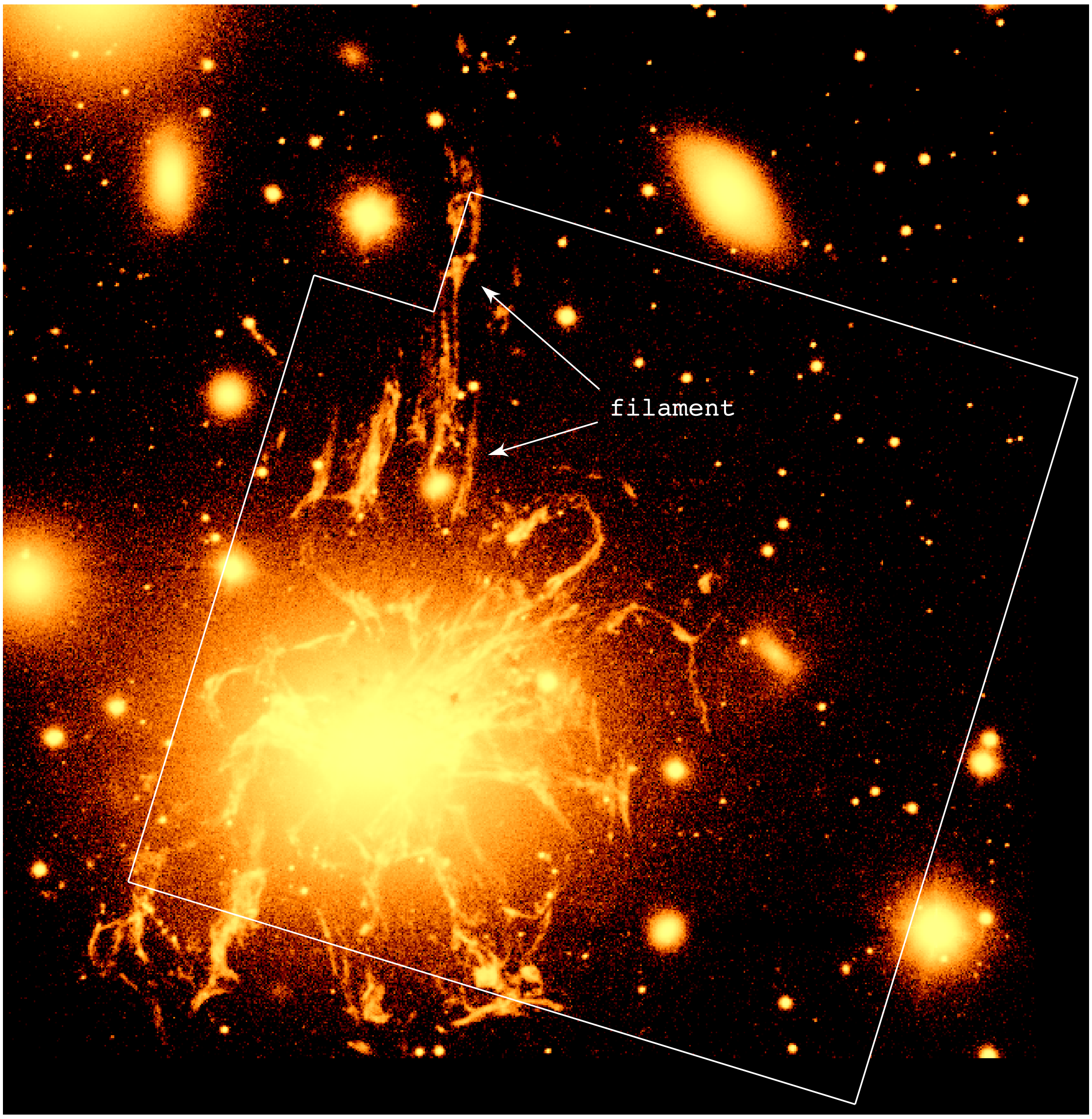

Fig E7. The SXS field is overlaid .on the cold gas nebulosity surrounding NGC1275. The image shows Hα emission[32]. The radial velocity along the long Northern filament measured from CO data[21] decreases, South to North (within the SXS field of view), from about +50 to -65 km/s. This is similar to the trend seen in the SXS velocity map (E6).

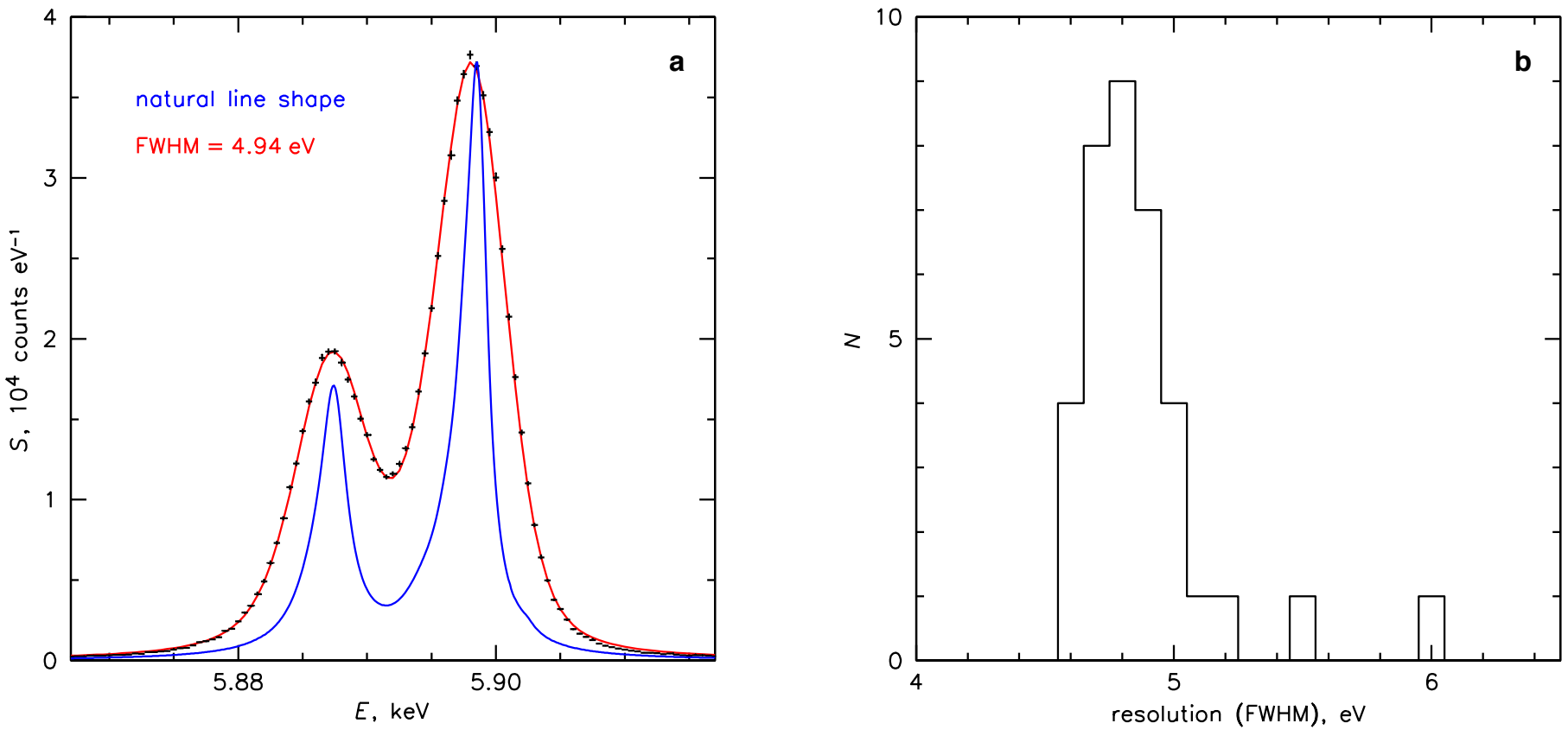

Fig E8. In-flight spectral resolution of the SXS. a) the composite spectrum of all pixels (excluding the cal. pixel) when they were exposed to the $^{55}Fe$ source on the filter wheel. b) the histogram of pixel resolution.

| Energy (eV) | λ (Å) | Charge state | Transition | Label | Note | Ref |
|---|---|---|---|---|---|---|
| **He-α** multiplet | | | | | | |
| 6617.00 | 1.8737 | | | | Blend – identified in (35) as Be- and Li-like iron | Fit |
| 6628.93 | 1.8704 | XXIII | $1s2s^22p\ ^1P_1 \rightarrow 1s^22s^2\ ^1S_0$ | | Be-like | 35 |
| 6636.84 | 1.8681 | XXV | $1s2s\ ^3S_1 \rightarrow 1s^2\ ^1S_0$ | He α (z) | Forbidden | 35 |
| 6645.24 | 1.8658 | XXIV | $1s2p^2\ ^2D_{5/2} \rightarrow 1s^22p\ ^2P_{3/2}$ | | Li-like | 35 |
| 6654.19 | 1.8633 | XXIV | $1s2s2p\ ^2P_{1/2} \rightarrow 1s^22s\ ^2S_{1/2}$<br>$1s2p^2\ ^2D_{3/2} \rightarrow 1s^22p\ ^2P_{1/2}$ | | Li-like blend | 35 |
| 6662.09 | 1.8610 | XXIV | $1s2s2p\ ^2P_{3/2} \rightarrow 1s^22s\ ^2S_{1/2}$ | | Li-like | 35 |
| 6667.90 | 1.8594 | XXV | $1s2p\ ^3P_1 \rightarrow 1s^2\ ^1S_0$ | He α (y) | Intercombination | 35 |
| 6682.45 | 1.8554 | XXV | $1s2p\ ^3P_2 \rightarrow 1s^2\ ^1S_0$ | He α (x) | Intercombination | 35 |
| 6700.76 | 1.8503 | XXV | $1s2p\ ^1P_1 \rightarrow 1s^2\ ^1S_0$ | He α (w) | Resonance | 35 |
| **H-like** doublet | | | | | | |
| 6951.96 | 1.7834 | XXVI | $2p\ ^2P_{1/2} \rightarrow 1s\ ^2S_{1/2}$ | Ly α2 | | 36 |
| 6973.18 | 1.7780 | XXVI | $2p\ ^2P_{3/2} \rightarrow 1s\ ^2S_{1/2}$ | Ly α1 | | 36 |
| **He-β** doublet | | | | | | |
| 7871.31 | 1.5751 | XXV | $1s3p\ ^3P_1 \rightarrow 1s^2\ ^1S_0$ | He β2 | | 37 |
| 7880.67 | 1.5733 | XXV | $1s3p\ ^1P_1 \rightarrow 1s^2\ ^1S_0$ | He β1 | | 37 |